\documentclass[aip,amsmath,amssymb,reprint
]{revtex4-1}

\usepackage{graphicx}
\usepackage{dcolumn}
\usepackage{bm}

\usepackage[utf8]{inputenc}
\usepackage[T1]{fontenc}
\usepackage{mathptmx}
\usepackage{etoolbox}

\makeatletter
\def\@email#1#2{
 \endgroup
 \patchcmd{\titleblock@produce}
  {\frontmatter@RRAPformat}
  {\frontmatter@RRAPformat{\produce@RRAP{*#1\href{mailto:#2}{#2}}}\frontmatter@RRAPformat}
  {}{}
}
\makeatother
\usepackage{csquotes}
\usepackage[colorlinks = true, linkcolor = blue, urlcolor = blue, citecolor = blue, anchorcolor = blue]{hyperref}
\usepackage{color}
\usepackage{bbm} 
\usepackage{amsfonts,amsmath,amssymb}
\usepackage{subfigure} 
\usepackage{mathrsfs}
\usepackage{verbatim}
\usepackage{centernot}
\usepackage{stackengine}

\usepackage{array}
\usepackage{bm}	
\renewcommand{\vec}[1]{\bm{#1}}

\begin{document}
\title{High-contrast double Bragg interferometry via detuning control}

\author{Rui Li}
\affiliation{Leibniz University Hanover, Institute of Quantum Optics, Hannover, Germany}
\author{V\'ictor Jos\'e Mart\'inez-Lahuerta}
\affiliation{Leibniz University Hanover, Institute of Quantum Optics, Hannover, Germany}
\author{Naceur Gaaloul}
\affiliation{Leibniz University Hanover, Institute of Quantum Optics, Hannover, Germany}
\author{Klemens Hammerer}
\affiliation{Leibniz University Hanover, Institute of Theoretical Physics, Hannover, Germany}
\affiliation{Institute for Theoretical Physics, University Innsbruck, 6020 Innsbruck, Austria}
\affiliation{Institute for Quantum Optics and Quantum Information, Austrian Academy of Sciences, 6020 Innsbruck, Austria}

\begin{abstract}
We propose high-contrast Mach-Zehnder atom interferometers based on double Bragg diffraction (DBD) operating under external acceleration. To mitigate differential Doppler shifts and experimental imperfections, we introduce a tri-frequency laser scheme with dynamic detuning control. We evaluate four detuning-control strategies—conventional DBD, constant detuning, linear detuning sweep (DS-DBD), and a hybrid protocol combining detuning sweep with optimal control theory (OCT)—using exact numerical simulations and a five-level S-matrix model. The OCT strategy provides the highest robustness, maintaining contrast above 95\% under realistic conditions, while the DS-DBD strategy sustains contrast above 90\% for well-collimated Bose-Einstein condensates. These results offer practical pathways to high-contrast, large-momentum-transfer DBD-based interferometers for precision quantum sensing and fundamental physics tests.
\end{abstract}
   

\date{\today}
\maketitle 
\section{Introduction}
Atom interferometry (AI) enables precision measurements of inertial and fundamental physical quantities by coherently splitting and recombining atomic wave packets along distinct paths. Applications range from atomic gravimetry~\cite{Peters-Nature-1999, Peters-Metrologia-2001, Poli-PRL-2011, Hu-PRA-2013, Altin-NJP-2013, Karcher-NPJ-2018, Szigeti-PRL-2020} and gravity gradiometry~\cite{Snadden-PRL-1998, McGuirk-PRA-2002, Rosi-PRL-2015}, rotation and inertial sensing~\cite{Gustavson-PRL-1997, Stockton-prl-2011, Geiger-NatCom-2011, Gautier-SciAdv-2022, Castanet-NatCom-2024, Stolzenberg-PRL-2025, Pelluet-NatCom-2025}, to precision determinations of the fundamental constants~\cite{Fixler-Science-2007, Lamporesi-PRL-2008, Rosi-Nature-2014, Muller-Science-2018, Morel-Nature-2020} and searches for new physics beyond the Standard Model such as ultralight dark matter~\cite{Geraci-PRL-2016, Arvanitaki-PRD-2018, Stadnik-Flambaum-PRA-2016, Graham-PRD-2016, Du-PRD-2022}.
A central challenge in advancing interferometer performance is achieving large momentum transfer (LMT) while preserving high contrast.

\begin{figure}[h]
  \centering
  \includegraphics[width=1.0\columnwidth]{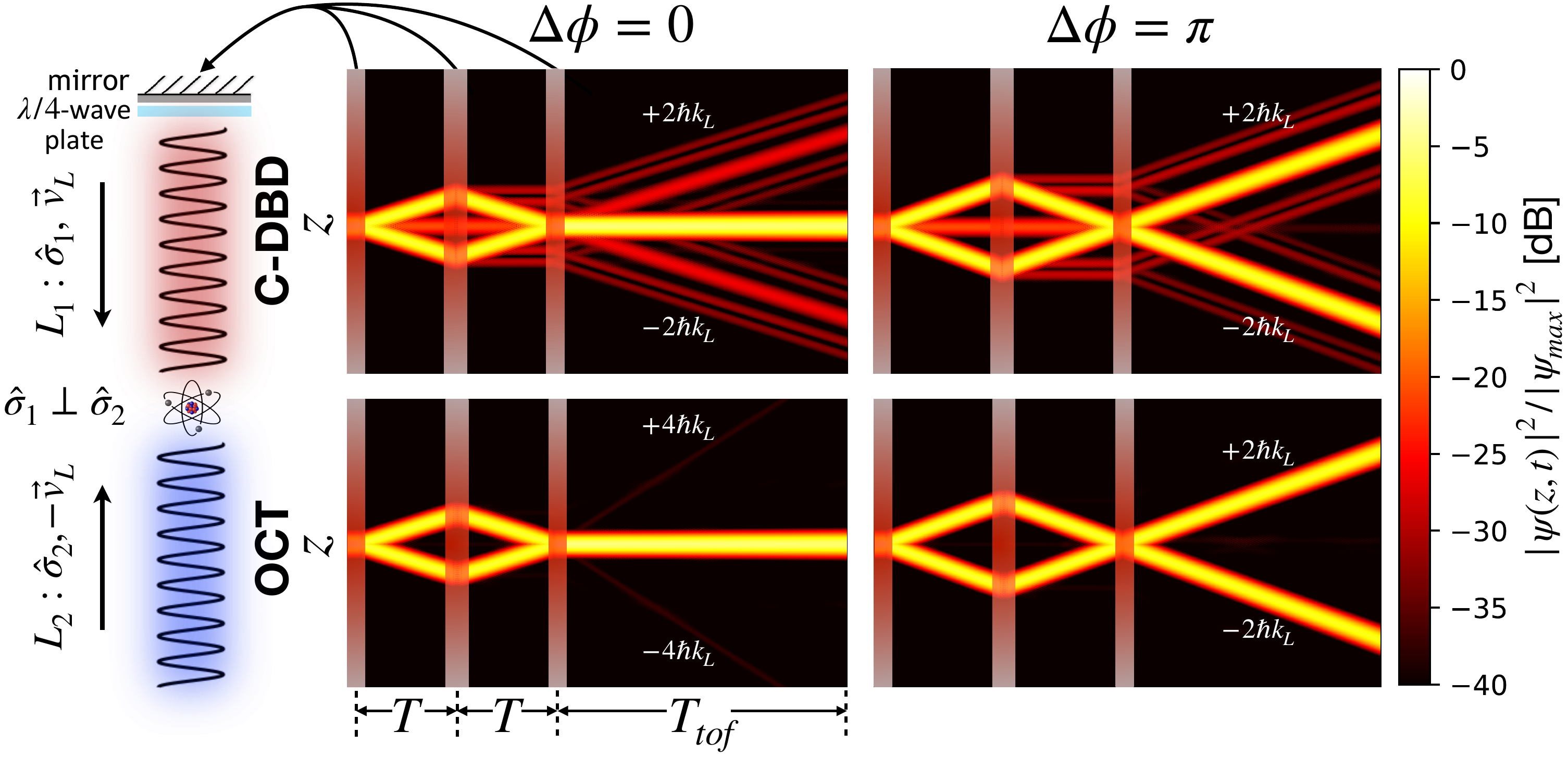}
    \caption{Schematics of a double Bragg atom interferometer under microgravity. Left: Experimental setup of a DBD pulse using counter-propagating optical lattices $L_1$ (red shaded) and $L_2$ (blue shaded) with orthogonal polarizations $\hat{\sigma}_1$ and $\hat{\sigma}_2$. Right: Atomic density evolution in the twin-lattice center-of-mass frame $|\psi(z,t)|^2$, normalized to its initial maximum $|\psi_{max}|^2=\max_{z}|\psi(z,0)|^2$ and shown in decibel units, for conventional (C-DBD) and optimized (OCT) Mach-Zehnder interferometers with phase shifts $\Delta \phi =0$ (left column) and $\pi$ (right column), adjusted via the interrogation time $T$. Atomic densities are obtained from exact numerical simulations, with red-shaded regions indicating the three DBD pulses. Only the visibly populated diffracted ports are labeled in the right panel after the last DBD pulse. 
    }
    \label{fig: AI}
 \end{figure}

Double Bragg diffraction (DBD) is an LMT technique used in atom interferometry that couples atoms to symmetric momentum states via two counter-propagating optical lattices with orthogonal polarizations (Fig.~\ref{fig: AI}, left panel), first demonstrated by the group of Ernst Maria Rasel and his collaborators within the QUANTUS consortium~\cite{Ahlers-PRL-2016}.
Compared to single Bragg and Raman schemes~\cite{Torii-PRA-2000, Bordé-PLA-1989, Kasevich-Chu-PRL-1991, Kasevich-Chu-APB-1992}, DBD doubles the interferometric scale factor at a given order while operating entirely within one internal state, eliminating hyperfine-state decoherence channels~\cite{Ozeri-2005-PRL, Uys-PRL-2010}. Relative to double Raman diffraction (DRD)~\cite{Gauguet-DRD-2009}, DBD is significantly simpler to implement experimentally: its required two-photon detuning typically lies in the kHz regime (rather than GHz), unwanted acousto-optic-modulator-generated sidebands are easily removed in free space, and the same laser system can drive subsequent Bloch oscillations (BO) for momentum transfer at the $10^2$–$10^3\,\hbar k_L$ scale~\cite{Gebbe-NatCom-2021}, where $k_L$ is the laser wave number. Moreover, higher-order single diffraction up to $24\,\hbar k_L$~\cite{Mueller-PRL-2008} and sequential diffractions up to $102\,\hbar k_L$~\cite{Chiow-prl-2011} have only been demonstrated with Bragg diffraction to date. DBD also exhibits an intrinsic parity symmetry~\cite{Li-PRR-2024} that suppresses common-mode phase noise, making it particularly suitable for horizontal geometries and microgravity platforms~\cite{Barrett-NatComm-2016, Becker-Nature-2018, Gaaloul-NatCom-2022, Elliott-Nature-2023}. However, the conventional retro-reflected DBD configuration~\cite{Ahlers-PRL-2016, Gebbe-NatCom-2021} cannot operate in non-degenerate settings (e.g., vertical geometries under gravity), which limits its competitiveness and practical applicability.

Interferometric contrast (or visibility) is a central figure of merit determining both the sensitivity and operational limits of atom interferometers. It determines the maximum usable interrogation time $T$ and effective momentum transfer $k_{\text{eff}}$, beyond which contrast loss dominates the measurement uncertainty. For a Mach–Zehnder (MZ) atom interferometer operating at the shot-noise limit, the acceleration sensitivity is given by
$\delta a = 1/(\mathbf{C}\sqrt{N}\, k_{\mathrm{eff}} T^{2})$,
where $N$ is the number of uncorrelated atoms detected and $\mathbf{C}$ is the contrast (or visibility)~\cite{Storey-Cohen-Tannoudji-1994, Schleich-Greenberger-Rasel-PRL-2013, Hardman-PRA-2014, Kritsotakis-Szigeti-PRA-2018}. While near-unity contrast can be achieved in SBD-based interferometers~\cite{Torii-PRA-2000}, DBD-based schemes are typically more susceptible to polarization errors, momentum spreads, and parasitic-path phase shifts~\cite{Hartmann-PRA-2020, Kirsten-Siemss-Fitzek-PRL-2023, Victor-arXiv-2025}, which reduce the efficiency of both the beam-splitter (BS) and—more severely—the mirror (M) pulses, ultimately limiting the achievable contrast in state-of-the-art DBD interferometers~\cite{Ahlers-PRL-2016, Gebbe-NatCom-2021, Stolzenberg-PRL-2025}.

In this work, we extended double Bragg diffraction to regimes involving large external accelerations—relevant for applications such as gravimetry—by introducing a third, dynamically tunable laser frequency that restores symmetry between the upward and downward Bragg transitions in Sec.~\ref{sec II}. We develop a microscopic description of the full DBD Mach–Zehnder interferometer using a five-level S-matrix formalism and compare the robustness of different detuning-control strategies in Sec.~\ref{sec III}. Building upon the detuning-sweep (or adiabatic passage) and optimal-control-theory (OCT) methods introduced in the previous works~\cite{Kovachy-ARP-PRA-2012, Li-PRR-2024, Peirce-PRA-1988, Riahi-PRA-2016, Saywell-PRA-2018, Saywell-JPB-2020, Louie-NJP-2023, Saywell-NatCom-2023, Goerz-Atoms-2023, Rodzinka-NatCom-2024, LeDesma-PRR-2024}, we systematically generalize these techniques to include the DBD mirror pulse—historically the main performance bottleneck. We find that OCT yields a mirror pulse achieving more than $99.5\%$ efficiency for an ultracold atomic ensemble with momentum width $0.05\,\hbar k_L$. Together, these developments provide a realistic and experimentally feasible route toward robust high-contrast DBD interferometers, with the potential to exceed $95\%$ overall contrast under realistic imperfections using advanced detuning-controlled pulses, expanding the applications of double Bragg diffraction in precision quantum sensing on both terrestrial and space-based platforms.

\section{Double Bragg accelerometers and detuning-control strategies} \label{sec II}
\subsection{Tri-frequency double Bragg accelerometers}
\begin{figure}[t]
  \centering
  \includegraphics[width=0.95\columnwidth]{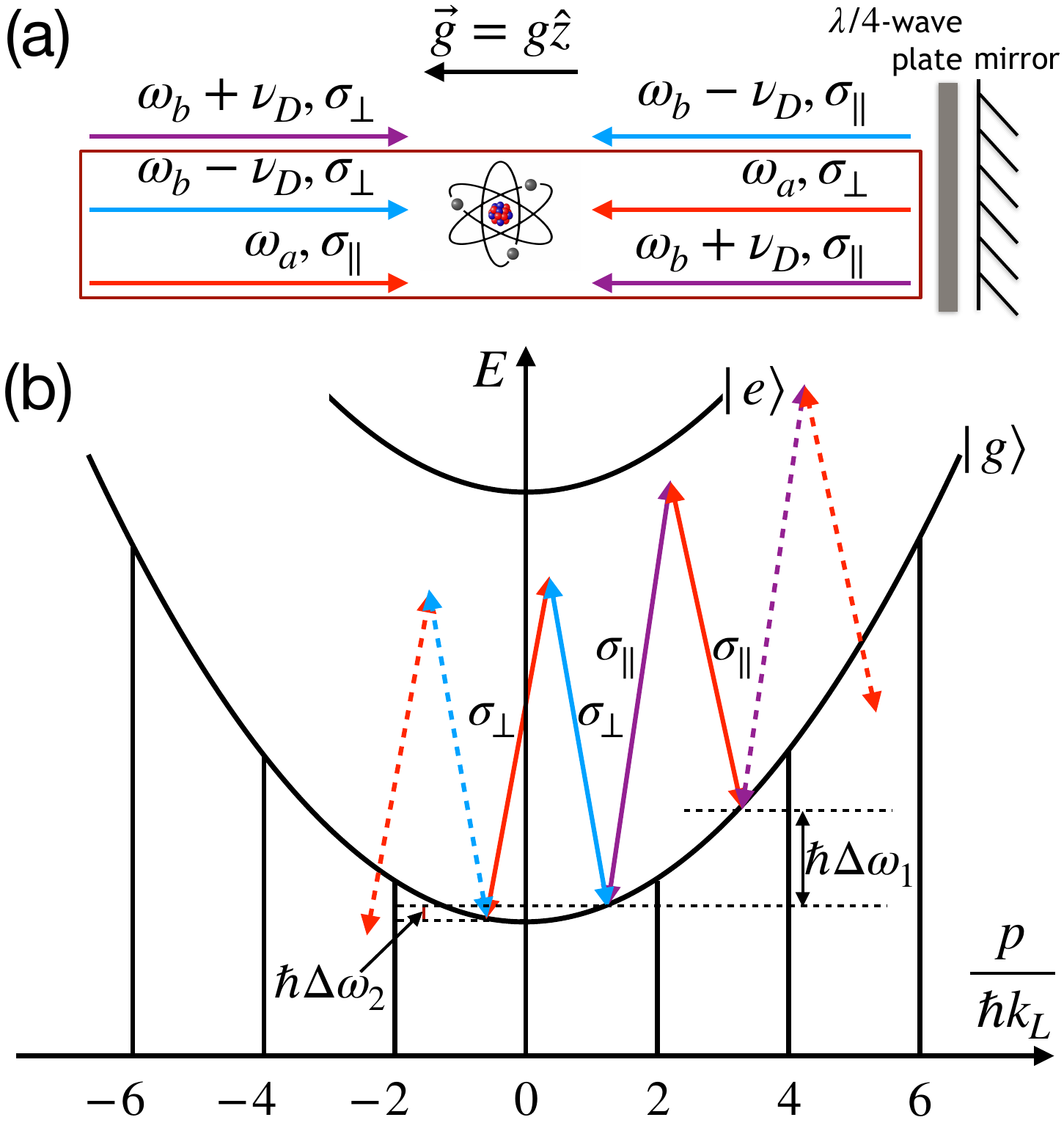}
    \caption{(a) Tri-frequency laser configuration enabling double Bragg diffraction of atoms under a strong constant acceleration $\vec{g} = g \hat{z}$, with dynamic Doppler shift compensation. The red box highlights four beams that resonantly drive DBD as the Doppler shift increases. (b) Corresponding energy-level diagram including Doppler and AC-Stark shifts. Upward and downward Bragg transitions are driven by frequency differences $\Delta \omega_{1,2} = \omega_b - \omega_a \pm \nu_D$, where $\nu_D$ is dynamically tuned to compensate the Doppler shift $\nu_g = 2k_L g t$. In (b), the solid lines indicate resonant transitions, whereas dashed lines represent off-resonant couplings.}
    \label{fig: DBD}
 \end{figure}

To measure a constant external acceleration $\vec{g}=g\hat z$, we consider a Mach–Zehnder atom interferometer based on DBD. This constant-acceleration assumption is fully compatible with real-world applications—such as gravimetry, geodesy, and inertial navigation—where the acceleration may vary between shots but is effectively constant over a single interferometer cycle. In microgravity or weak-acceleration regimes, a standard dual-frequency scheme suffices to drive symmetric momentum transfer in opposite directions~\cite{Gebbe-NatCom-2021, Ahlers-PRL-2016}. However, under strong acceleration—such as gravity—the resulting differential Doppler shift $\nu_g = 2\,k_L g t$ (frequencies are given in angular units unless stated
otherwise) breaks this symmetry, preventing simultaneous resonance with the upward and downward Bragg transitions. To overcome this limitation, we propose a tri-frequency retro-reflective configuration (see Fig.~\ref{fig: DBD}(a)), previously demonstrated in double Raman gravimeters~\cite{Malossi-PRA-2010}. In this configuration, one of the two input frequencies (e.g., $\omega_b$) is replaced by a frequency pair $\omega_b \pm \nu_D$, where the detuning $\nu_D = 2k_L a_L t$ linearly compensates the Doppler shift $\nu_g$. Due to momentum selectivity~\cite{Li-PRR-2024}, the accelerating atoms couple only to four of the six beams, forming a pair of resonant Bragg lattices highlighted in Fig.~\ref{fig: DBD}(a).

The one-dimensional (1D) single-particle Hamiltonian describing tri-frequency double Bragg diffraction of an atom with mass $m$ subjected to a strong constant acceleration $\vec{g} = g \hat{z}$, as illustrated in Fig.~\ref{fig: DBD}(a–b), can be written in the laboratory frame as
\begin{align}
    H_{lab}(t)=&\frac{\hat p^2}{2m} +2\hbar \Omega(t) \cos\Big(2 k_L\hat z-\int \!\nu_D(t) \,dt\Big)\times \nonumber\\
    &\Big\{\cos\big[\Delta \omega (t)t\big]+\varepsilon_{pol}\Big\}-mg\hat z,
    \label{Eq: H_lab}
\end{align}
where $g>0$ and $\varepsilon_{pol}=|\mathbf{\sigma_\perp}^\dagger\mathbf{\sigma_{\parallel}}|$ quantifies the polarization error arising from imperfect optical-field polarizations. For a conservative (worst-case) estimate of polarization-error effects, we neglect the temporal modulation associated with this term. A detailed derivation of Eq.~\eqref{Eq: H_lab} is provided in the Supplementary Material. $\Omega(t)$ is an effective two-photon Rabi frequency for the light-atom interaction, which can be time-dependent. For an MZ atom interferometer with triple Gaussian pulses (see Fig.~\ref{fig: AI}, right panel), $\Omega(t)$ takes the following form:
\begin{align}
    \Omega(t) = \Omega_{BS} e^{-\frac{t^2}{2\tau_{BS}^2}} + \Omega_{M} e^{-\frac{(t-T)^2}{2\tau_{M}^2}} + \Omega_{BS} e^{-\frac{(t-2T)^2}{2\tau_{BS}^2}}, \label{Eq: Rabi}
\end{align}
where the three light pulses are centered at time ($0$, $T$, $2T$) with peak two-photon Rabi frequencies ($\Omega_{BS}$, $\Omega_{M}$, $\Omega_{BS}$) and pulse widths ($\tau_{BS}$, $\tau_{M}$, $\tau_{BS}$), respectively. 
It should be noted that
$\Delta \omega (t) t = 4\,\omega_{\mathrm{rec}} t + \Delta(t) t \equiv \phi (t_i) + \int_{t_i}^{t} [\omega_b(t)-\omega_a(t)] \, dt $ in Eq.~\eqref{Eq: H_lab}, with $\omega_{\mathrm{rec}}\equiv\hbar k_L^2/(2 m)$, denotes the physically accumulated phase difference between the blue (or purple) and red lasers shown in Fig.~\ref{fig: DBD}(a-b), evaluated in the center-of-mass (COM) frame of the twin Bragg lattices where $\nu_D=0$. In this COM frame, the Hamiltonian in Eq.~\eqref{Eq: H_lab} transforms into 
\begin{align}
    H_{COM}(t)=&\frac{\hat p^2}{2m} +2\hbar \Omega(t) \cos[2 k_L\hat z]\Big\{\cos\big[\Delta \omega (t)t\big]+\varepsilon_{pol}\Big\} \nonumber\\&-m(g-a_L)\hat z 
    \label{Eq: H_lattice}
\end{align}
via a time-dependent unitary transformation $H_{COM}(t) = U H_{lab}(t) U^\dagger +i\hbar \dot U U^\dagger $ (see Supplementary Material for details). $a_L=\dot \nu_D/(2k_L)$ is the COM acceleration of the twin Bragg lattices which can be fine-tuned to approach the linear acceleration $g$ such that $g-a_L$ is approaching zero. From Eq.~\eqref{Eq: H_lattice}, one identifies an effective acceleration in the transformed frame, $g_{eff}\equiv g-a_L$. For the remainder of this paper, we refer to $g_{eff}$ simply as $g$, and perform all calculations in the twin-lattice COM frame unless otherwise specified.

Under the microgravity assumption in the transformed frame, i.e., $|m g z| \ll 2\hbar\Omega_R $ and $|g| \ll \hbar k_L/(m t)$ for all relevant times $t$ and positions $z$ during the interferometer sequence, the last term in Hamiltonian~\eqref{Eq: H_lattice} can be treated as a perturbation and neglected during the light-matter interaction. This allows the application of the double Bragg theory developed in Ref.~\cite{Li-PRR-2024}. During free evolution intervals between pulses, the effective acceleration leads to both a momentum shift of the atomic wave packet and an additional propagation phase dependent on the initial momentum state $|p\rangle$ and the free evolution time $T$:
\begin{align}
    \hat{U}(T)|p\rangle &= e^{-\frac{i T}{\hbar } \big( \frac{\hat p^2}{2m}-m g \hat z \big)}|p\rangle \nonumber \\&= \exp{\left(-\frac{i }{2 m \hbar}(mg T^2 p +T p^2)\right)}|p+m g T\rangle \nonumber \\&\equiv U(p)|p+m g T\rangle ,\label{Eq:propagation phase}
\end{align}
where the global phase proportional to $T^3$ has been neglected~\cite{Kritsotakis-Szigeti-PRA-2018}. 
Since the propagation phase only depends on the momentum state $|p\rangle$, the spatially parallel trajectories will accumulate the same phase during the free fall in between successive pulses. Every applied DBD pulse will split each trajectory into different momentum classes, and hence, resulting in a coherent superposition of phase contributions from all intermediate trajectories at the final detection port.
\subsection{Detuning-control strategies for high-contrast double Bragg interferometers}
We now lay out four detuning-control strategies aimed at maximizing beam-splitter and mirror pulse efficiencies for Doppler-broadened wave packets with finite momentum width. Detailed pulse parameters, detuning profiles, and analyses of single-pulse efficiency and robustness are provided in the online Supplementary Material.
\subsubsection{Conventional-DBD Mach-Zehnder protocol}
\begin{figure}[t]
  \centering
  \includegraphics[width=1.0\columnwidth]{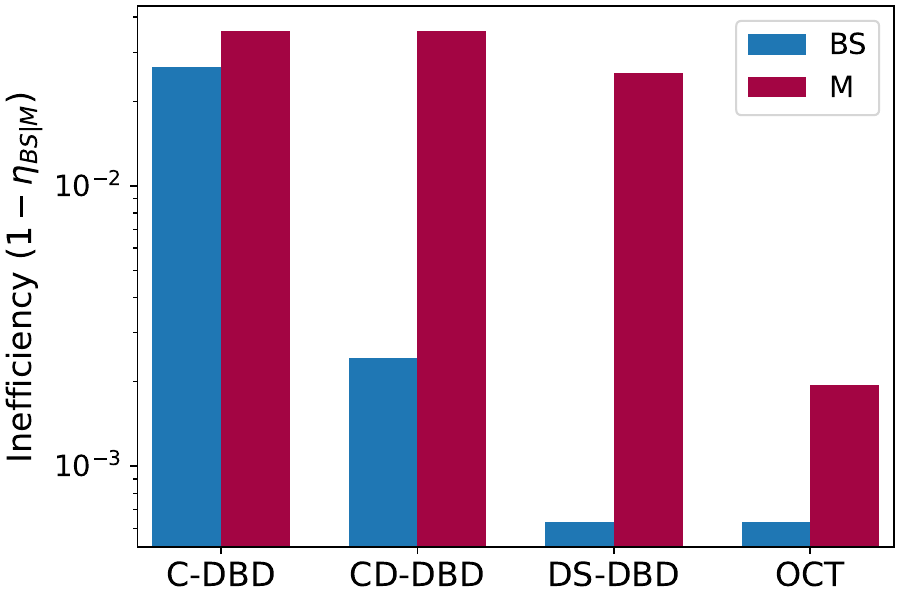}
    \caption{Comparison of beam-splitter (BS) and mirror (M) inefficiencies ($1-\eta_{BS| M}$) for four strategies: conventional DBD (C-DBD), constant-detuning DBD (CD-DBD), linear-detuning-sweep DBD (DS-DBD), and a hybrid strategy combining DS-BS with OCT-M (OCT). BS and M efficiencies $\eta_{BS| M}$ are evaluated for an input Gaussian wave packet with a momentum width $\sigma_p=0.05\,\hbar k_L$, centered at $p_0 = 0$ for BS and $p_0 = 2\,\hbar k_L$ for M without polarization error ($\varepsilon_{pol}=0$).
    }
    \label{fig: efficiency_comparison}
 \end{figure}

We begin with the Mach-Zehnder sequence based on conventional double Bragg diffraction (C-DBD), where both beam-splitter and mirror pulses operate at the standard first-order DBD resonance condition $\Delta \omega= 4\,\omega_{\mathrm{rec}}$ (or $\Delta(t)=0$)~\cite{Giese-PRA-2013, Giese-FdP-2015, Ahlers-PRL-2016, Li-PRR-2024}. The Gaussian pulse widths are individually optimized at fixed peak Rabi frequencies to balance momentum acceptance window and suppress higher-order diffraction losses at the same time. We choose $\Omega_{BS} = 2.0\,\omega_{\mathrm{rec}}$ and $\Omega_{M} = 2.89\,\omega_{\mathrm{rec}}$, with corresponding optimal widths $\tau_{BS} = 0.47\,\omega_{\mathrm{rec}}^{-1}$ and $\tau_{M} = 0.64\,\omega_{\mathrm{rec}}^{-1}$. For an input Gaussian wave packet with $\sigma_p = 0.05\,\hbar k_L$ centered at $p_0 = 0$ (for BS) and $2\,\hbar k_L$ (for M), the efficiencies are $\eta_{BS} = 97.348\%$ and $\eta_{M} = 96.426\%$, respectively (see Fig.~\ref{fig: efficiency_comparison}). The full C-DBD sequence consists of a BS–M–BS configuration with interrogation time T between pulses and serves as our baseline for comparison.
\subsubsection{Constant-detuning mitigated Mach-Zehnder protocol}
The second strategy employs constant detuning mitigation, where the beam-splitter and mirror pulses of the C-DBD scheme are improved by adding a fixed detuning to compensate known polarization errors and AC-Stark shift~\cite{Li-PRR-2024}. This yields a modified resonance condition: $\Delta \omega= 4\,\omega_{\mathrm{rec}}+ \Delta$ with $\Delta = \text{Const.}$. Optimal detunings are found to be $\Delta_{BS} = 0.27\,\omega_{\mathrm{rec}}$ for the BS and $\Delta_{M} = 0$ for the mirror, using the same Gaussian pulses as in C-DBD. Without polarization error ($\varepsilon_{pol}=0$), the BS and mirror pulse efficiencies are $\eta_{BS} = 99.757\%$ and $\eta_{M} = 96.426\%$, respectively, for an input Gaussian wave packet with $\sigma_p=0.05\,\hbar k_L$ centered at $p_0=0$ and $2\,\hbar k_L$ (see Fig.~\ref{fig: efficiency_comparison}). The resulting constant-detuning DBD (CD-DBD) sequence follows a BS–M–BS layout with fixed interrogation time T. This approach is effective only when polarization errors are known and the momentum distribution is narrow, ideally using box-shaped pulses.
\subsubsection{Linear-detuning-sweep mitigated Mach-Zehnder protocol}
The third strategy further improves the C-DBD and CD-DBD protocols by applying time-dependent linear detuning sweeps to mitigate AC-Stark and Doppler shifts, thereby improving beam-splitter and mirror pulse efficiencies~\cite{Li-PRR-2024}. The detuning follows $\Delta(t)/\omega_{\mathrm{rec}} = (\alpha/\tau_{BS|M})(t - t_0)+\beta$ where $t_0$ denotes the center of the respective Gaussian pulse, with optimized parameters $(\alpha_{BS}, \beta_{BS}) = (0.37, 0.315)$ and $(\alpha_{M}, \beta_{M}) = (0.75, -4)$, using the same Gaussian pulses as in C-DBD.

\begin{figure}[t]
  \centering
  \includegraphics[width=1.0\columnwidth]{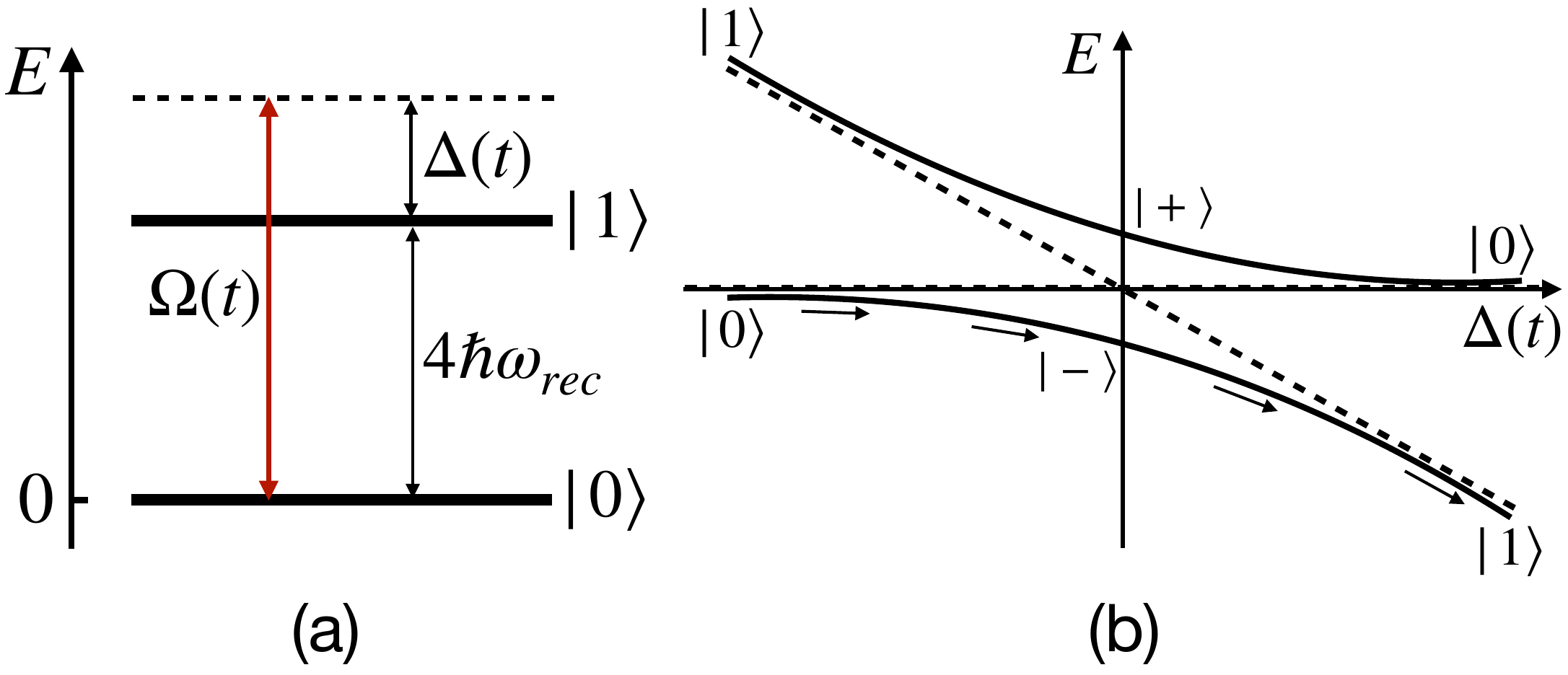}
    \caption{(a) Effective two-level system representing first-order DBD developed in Ref.~\cite{Li-PRR-2024}. (b) Linear detuning sweep mimicking adiabatic passage for robust population transfer. The initial and final states of an ideal double Bragg beam-splitter are given by $|0\rangle=|0\hbar k_L\rangle$ and $|1\rangle=(|2\hbar k_L\rangle + |-2\hbar k_L\rangle)/\sqrt{2}$, respectively.
    }
    \label{fig: detuning-sweep DBD}
 \end{figure}

This approach addresses the momentum-dependent energy shift arising from time-varying AC-Stark shifts, Doppler effects due to finite $\sigma_p$ or nonzero $p_0$, and polarization errors $\varepsilon_{pol}$—all of which may fluctuate shot-to-shot and render the resonance condition analytically intractable. To circumvent the need for precise resonance knowledge, we adopt a detuning control strategy inspired by the principle of \textit{adiabatic passage} in two-level systems~\cite{Steck-Quantum-Optics, Landau-PJSU-1932, Zener-PRSL-1932}, a well-established technique in nuclear magnetic resonance~\cite{Bloch-Siegert-PhysRev-1940, Bloch-physrev-1946, Abragam-1961-Oxford, Baum-PRA-1985, Philpott-1987}. We extend this concept to the multi-level case of double Bragg diffraction via a Magnus expansion in the quasi-Bragg regime~\cite{Li-PRR-2024}, yielding an effective two-level model between $|0\rangle = |0\hbar k_L\rangle$ and $|1\rangle = (|2\hbar k_L\rangle + |-2\hbar k_L\rangle)/\sqrt{2}$ (see Fig.~\ref{fig: detuning-sweep DBD}(a)).

A linear detuning sweep enables robust and highly efficient population transfer (illustrated in Fig.~\ref{fig: detuning-sweep DBD}(b)), yielding a beam-splitter efficiency of $\eta_{BS} = 99.937\%$ at $\varepsilon_{pol} = 0$. This represents a substantial improvement over both C-DBD and CD-DBD, with a residual inefficiency below $10^{-3}$ (see Fig.~\ref{fig: efficiency_comparison}). In contrast, the DS mirror offers only a modest gain of about $1\%$, achieving an efficiency of $\eta_M = 97.465\%$. This indicates that substantial room for further optimization via OCT remains. Alternative monotonic detuning profiles (e.g., sigmoid) were also explored but offered no significant gain. The resulting DS-DBD interferometer follows a BS–M–BS sequence with interrogation time $T$. This approach is effective when the range of polarization errors is known (e.g., $\varepsilon_{pol}<5\%$), and the uncertainties in $p_0$ and $\sigma_p$ are small and under control (e.g., for well-collimated BECs).
\subsubsection{Mach-Zehnder protocol with optimal control theory}
The final detuning-controlled Mach-Zehnder strategy targets the optimization of the mirror pulse—identified as the performance bottleneck in the DS-DBD protocol—by applying optimal control theory (OCT), implemented with QCTRL’s Boulder Opal package~\cite{QCtrlpackage}. Unlike SBD mirror pulse, the DBD mirror pulse involves a four-photon transition $|\pm 2\hbar k_L\rangle\rightarrow|0\hbar k_L\rangle\rightarrow|\mp 2\hbar k_L\rangle $ making it more sensitive to finite momentum spread and nonzero COM momentum~\cite{Szigeti_2012, Mueller-PRL-2008, Chiow-prl-2011, Stolzenberg-PRL-2025}. As illustrated in Fig.~\ref{fig: AI}, a C-DBD mirror selectively filters and only reflects the central momentum component. This sensitivity is mitigated by jointly optimizing a smooth time-dependent detuning $\Delta(t)$ and all three Gaussian pulse parameters $(\Omega_M,\, \tau_M,\, t_0)$, yielding an OCT mirror pulse with $(\Omega_M,\, \tau_M,\, t_0) = (2.502\, \omega_{\mathrm{rec}},\, 1.829\, \omega_{\mathrm{rec}}^{-1},\, 3.879\, \omega_{\mathrm{rec}}^{-1})$ and an efficiency of $\eta_{M} = 99.806\%$ (see Fig.~S8 in Supplementary Material for the OCT detuning $\Delta(t)$), representing a substantial improvement over the DS mirror pulse (see Fig.~\ref{fig: efficiency_comparison}). The resulting hybrid “OCT” sequence combines this mirror with two DS beam-splitters, separated by interrogation time $T$.

We also explored a fully OCT-optimized protocol, where both beam-splitter and mirror pulses are designed via OCT. However, no significant contrast improvement is observed over the DS-BS combined with OCT-mirror scheme under typical conditions with well-collimated BECs and low polarization error. A fully OCT-based approach becomes favorable only when polarization errors exceed 5\%, or when large uncertainties in $p_0$, $\sigma_p$, and $\varepsilon_{pol}$ occur simultaneously.
\section{Contrast of double Bragg Mach-Zehnder interferometers}\label{sec III}
For spatially unresolved interferometers, i.e., the population detection after the time-of-flight ($T_{tof}$) cannot distinguish parallel spatial trajectories, either due to a short interrogation time $T$ or because the wave packet expansion during the time-of-flight $T_{tof}$ is comparable to their spatial separation, the output ports essentially correspond to different momentum states (see Fig.~\ref{fig: AI}). In the far-field limit of $T_{tof}\gg T$, this correspondence becomes exact. In this case, the full interferometer can be fully described in momentum space, with its total S-matrix given by the time-ordered product of the S-matrices for individual pulses, interleaved with diagonal unitary matrices that encode the propagation phases of each momentum state accumulated between adjacent pulses. 
\begin{figure*}[!t]
  \centering
  \includegraphics[width=1.0\textwidth]{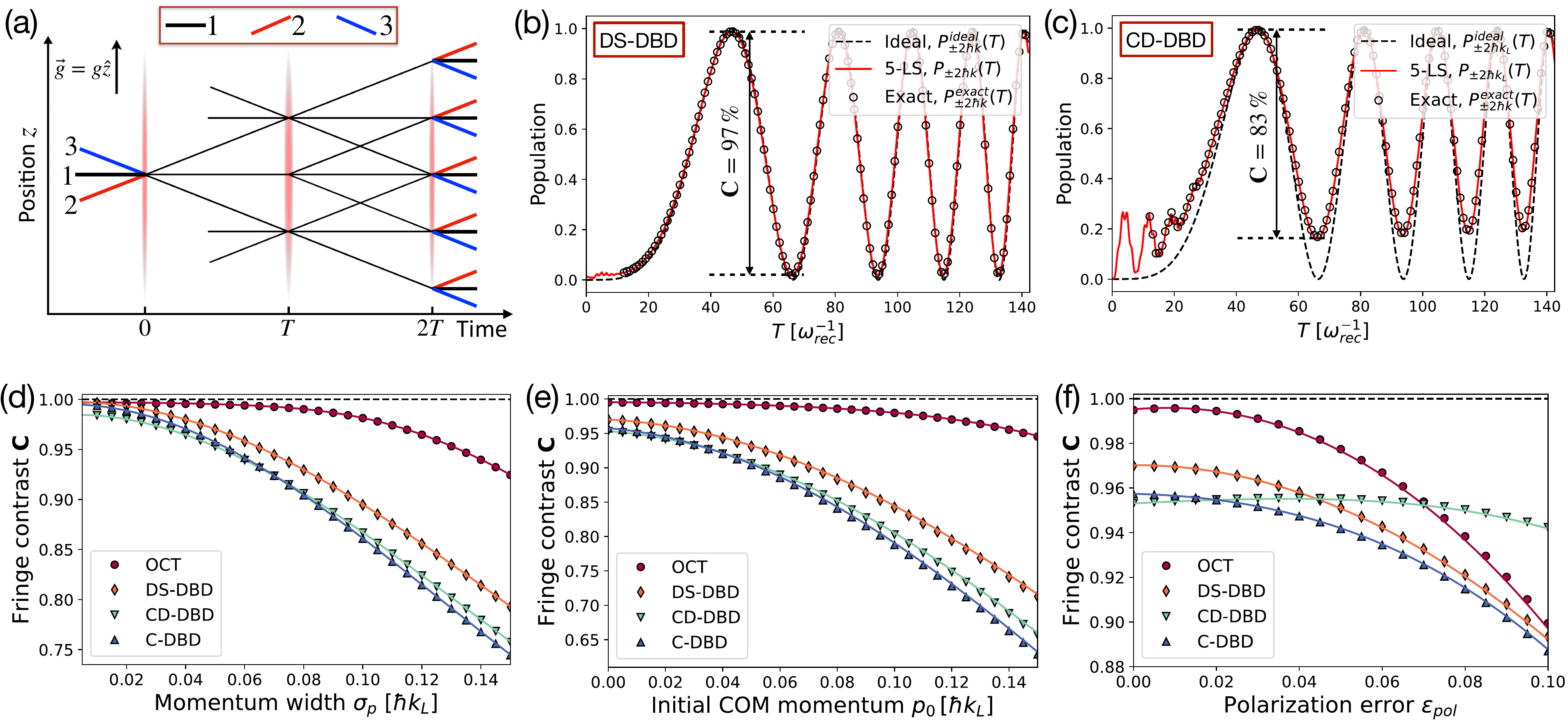}
    \caption{(a) Mach-Zehnder atom interferometer in momentum space implemented with three DBD pulses. Input and output ports are labeled by indices $i=1,2,3$, corresponding to momentum states $|p\rangle$, $|p+2\hbar k_L\rangle$, and $|p-2\hbar k_L\rangle$, respectively. Higher-order momentum states (e.g., $|\pm 4\hbar k_L\rangle$) are not shown for clarity but are included in the population calculations. (b) T-scan fringe in the $|\pm 2\hbar k_L\rangle $-port using the DS-DBD strategy shows 97\% contrast under an effective acceleration $ g = 0.000357 k_L^{-1}\omega_{\mathrm{rec}}^{2}$ with an initial momentum width $\sigma_p=0.05\hbar k_L$ and COM momentum $p_0=0$. (c) Contrast degradation in the CD-DBD protocol due to a large initial COM momentum $p_0=0.1\hbar k_L$ and a momentum width $\sigma_p=0.01\hbar k_L$. (d) Contrast versus momentum width $\sigma_p$ with vanishing COM momentum and polarization error. (e) Contrast versus initial COM momentum $p_0$ with a momentum width $\sigma_p=0.05\hbar k_L$ and no polarization error. (f) Contrast versus polarization error $\varepsilon_{pol}$ with an initial momentum width $\sigma_p=0.05\hbar k_L$ and vanishing COM momentum. In subplots (d–f), exact numerical results are shown as symbols, while solid curves represent predictions from the five-level S-matrix theory.
    }
    \label{fig: MZAI}
 \end{figure*}
The resulting S-matrix of the double Bragg Mach–Zehnder interferometer is therefore
\begin{align}
    S^{tot}=S^{BS}U(2T, T)S^{M}U(T, 0)S^{BS},\label{Eq: S_total}
\end{align}
where the contributions from all parasitic trajectories up to $\pm 4\hbar k_L$ momentum transfer are considered. The S-matrix of the beam-splitter pulse $S^{BS}$ in the ordered basis $\{|p\rangle,\,|p+2\hbar k_L\rangle,\,|p-2\hbar k_L\rangle,\,|p+4\hbar k_L\rangle,\,|p-4\hbar k_L\rangle \}$ is given by a $5\times5$ matrix $(B_{ij}(p))$ with $i,\,j=1,\dots,5$, and that of the mirror pulse $S^{M}$ is given by a matrix $(M_{ij}(p))$, both of which depend on the quasi-momentum $p$ and can only be solved numerically for a Gaussian pulse with a time-dependent detuning~\cite{Li-PRR-2024}. The two unitary matrices in above equation, given analytically by $U(T, 0)=(U_{ij}(p))$ and $U(2T, T)=(U_{ij}(p+mgT))$ with $(U_{ij}(p))= diag[U(p), U(p+2\hbar k_L), U(p-2\hbar k_L), U(p+4\hbar k_L),U(p-4\hbar k_L)]$ (see Eq.~\eqref{Eq:propagation phase} for $U(p)$), contain the propagation phase depending on effective acceleration $g$, interrogation time $T$ as well as the input quasi-momentum $p$. With above notations, an arbitrary S-matrix element of the full interferometer can be explicitly expressed as
\begin{align}
    S^{tot}_{ij}(g,p,T)=\sum_{k,l=1}^5 &B_{il}(p_3)U_{ll}(p_2)M_{lk}(p_2)U_{kk}(p_1)\times\nonumber\\& B_{kj}(p_1),\label{Eq: S_total_explicit}
\end{align}
where $p_1=p$, $p_2=p+m g T$ and $p_3=p+2m g T$. We choose to use a five-level S-matrix (5-LS) description of the double Bragg pulses in order to accurately capture the dynamics in the quasi-Bragg regime~\cite{Li-PRR-2024} even though most population stays within the first three Brillouin zones $[-3\hbar k_L,\,3\hbar k_L]$ during the full Mach-Zehnder interferometer for a quasi-momentum $p\ll \hbar k_L$. The final output state $|\psi^{out}\rangle$ after the full MZ sequence with an initial state of the form $ |\psi(t=0)\rangle=\int\! dp\, \psi(p)|p\rangle $ with $|\psi(p)|^2$ being a normalized momentum distribution with compact support in the first Brillouin zone $[-\hbar k_L,\, \hbar k_L]$, such as a Gaussian $\mathcal{N}(p_0,\sigma_p^2)$ with $p_0, \,\sigma_p\ll \hbar k_L$, is given by 
\begin{align}
    |\psi^{out}\rangle&= \int\! dp\, \psi(p)S^{tot}|p\rangle, 
\end{align}
where the final output basis is shifted due to effective acceleration to $\{|p_3\rangle,|p_3+2\hbar k_L\rangle,|p_3 -2\hbar k_L\rangle, |p_3 +4\hbar k_L\rangle, |p_3 -4\hbar k_L\rangle\}$. 
Therefore, the output wave functions for the three main detection ports—corresponding to the signals in the black, red, and blue output ports in Fig.~\ref{fig: MZAI}(a)—are given by $\phi_i(p)=\psi(p)S^{tot}_{i1}(g,p,T)$ for $i=1,\,2,\,3$. The integrated population in each detection port $i$ is then given by
\begin{align}
    P_i(g, T)&=\int_{-\hbar k_L}^{\hbar k_L} \left|\psi(p)\,S^{tot}_{i1}(g,p,T)\right|^2 \,dp,
\end{align}
respectively. Furthermore, for double Bragg interferometers, the populations in port 2 and 3 are summed to produce a signal conjugate to that in port 1, which, under ideal BS and mirror operations, takes the form of a single sinusoidal function~\cite{Sven_Abend_PhD_thesis}:
\begin{align}
     P_{\pm 2\hbar k}^{ideal}(g, T) = \frac{A-\mathcal{C}\cos[4k_L g T^2]}{2},\label{Eq: ideal fringe}
\end{align}
with an offset of $A=1$ and unity contrast ($\mathcal{C}=1$) , shown as a dashed black line in Fig.~\ref{fig: MZAI}(b–c).

For non-ideal beam-splitter or mirror operations, the contrast is generally less than unity, and the output signal typically contains multiple Fourier components, e.g., Ramsey–Bordé–type parasitic paths produce distinct oscillatory components from those of the main DBD interferometer~\cite{Kirsten-Siemss-Fitzek-PRL-2023}. However, the contrast can still be defined, analogous to the ideal case, as the population difference between the first two non-trivial extrema of the output signal when scanning the interrogation time $T$: 
\begin{align}
    \mathbf{C}\equiv  P_{\pm 2\hbar k}(g, T_{max}) - P_{\pm 2\hbar k}(g, T_{min}),
\end{align}
where $T_{max}$ and $T_{min}$ denote the first non-trivial maximum and minimum of $P_{\pm 2\hbar k}(g, T)\equiv P_{2}(g, T) + P_{3}(g, T) $ at a fixed $g$ as illustrated by the contrast extraction in Fig.~\ref{fig: MZAI}(b–c). For instance, Fig.~\ref{fig: MZAI}(b) shows that the contrast extracted for the DS-DBD strategy for an initial input state with a momentum width of $\sigma_p=0.05\hbar k_L$ and COM momentum $p_0=0$ without polarization error is $\mathbf{C}=97\%$. Fig.~\ref{fig: MZAI}(c) shows the detrimental effect of a large initial COM momentum, $p_0=0.1\hbar k_L$, on the contrast, where the contrast is reduced to $\mathbf{C}=83\%$ for a narrower momentum width of $\sigma_p=0.01\hbar k_L$ without polarization error. In both cases, the results of the five-level S-matrix theory agrees perfectly with the exact numerical solutions of the Schrödinger equation based on the second-order Suzuki-Trotter decomposition~\cite{Suzuki-Trotter-1990, Fitzek-UATIS-2020, Seckmeyer-FFTarray} with a momentum truncation up to $\pm 14.5\hbar k_L$. The single Fourier fringe model (Eq.~\eqref{Eq: ideal fringe}) applies only in the ideal regime of small initial COM momentum and momentum width, where pulse efficiency remains nearly unaffected.

\textit{Main results.} We compare the contrast robustness of different detuning-control strategies with respect to three dominant experimental imperfections: the atomic momentum width $\sigma_p$ [Fig.~\ref{fig: MZAI}(d)], the initial center-of-mass momentum $p_0$ [Fig.~\ref{fig: MZAI}(e)], and the polarization error $\varepsilon_{pol}$ [Fig.~\ref{fig: MZAI}(f)]. For realistic experimental conditions with well-controlled polarization errors ($\varepsilon_{pol} < 3\%$), we consistently observe the performance hierarchy
\[
\text{OCT} > \text{DS-DBD} > \text{C-DBD} \approx \text{CD-DBD}.
\] 
The relative improvements of the DS-DBD and OCT protocols over C-DBD (or CD-DBD) depend on the atomic momentum width (set by the 1D effective temperature), the COM momentum (influenced by laser alignment and pulse timing), and the polarization properties of the optical beams. 
For a typical atom-chip-generated BEC with a 1D effective temperature of $2 \,\text{nK}$~\cite{Deppner-PRL-38pK-2021}, corresponding to a momentum width of approximately $\sigma_p = 0.10\hbar k_L$ (for $^{87}\text{Rb}$ atoms at a wavelength of $780\,\text{nm}$), the DS-DBD and OCT strategies yield relative contrast improvements of $3.4\%$ and $12.0\%$, respectively, over the C-DBD protocol (assuming perfect polarizations and zero COM momentum). For a BEC after evaporative cooling with a 1D temperature below $500 \,\text{pK}$~\cite{Leanhardt-Science-2003}, or a momentum width about $ 0.05\hbar k_L$, the improvements reduce to $1.3\%$ (for DS-DBD) and $3.8\%$ (for OCT). In the state-of-the-art case of delta-kick-collimated BECs with an effective temperature below $40 \,\text{pK}$~\cite{Deppner-PRL-38pK-2021, Kovachy-PRL-2015}, corresponding to $\sigma_p = 0.014\hbar k_L$, the improvements are $0.4\%$ and $0.5\%$, respectively. With well-controlled polarization error and initial COM momentum, the DS-DBD strategy maintains contrast above $90\%$ up to $\sigma_p \leq 0.097\,\hbar k_L$, while the OCT strategy remains above $95\%$ contrast up to $\sigma_p \leq 0.132\,\hbar k_L$.

Finally, we investigate the robustness of the contrast against fluctuations in peak lattice depth for different detuning-control strategies. 
\begin{figure}[t]
    \centering
    \includegraphics[width=1.0\columnwidth]{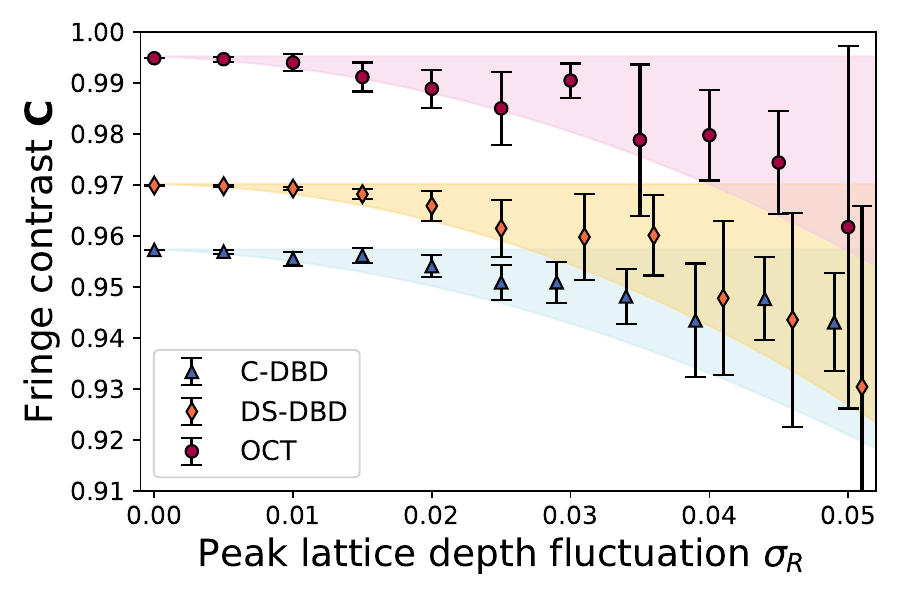}
    \caption{Contrast as a function of relative peak lattice depth fluctuations for different detuning-control strategies. Symbols with error bars (representing standard deviation) stand for exact numerical simulations; shaded regions denote theoretical confidence intervals from the five-level S-matrix model.}
    \label{fig: Robustness_against_peak_lattice_depth}
\end{figure}
Since all protocols are tuned to individually optimized parameters—including the peak lattice depth $\Omega_{BS|M}$, which is proportional to the peak laser power—they are expected to be first-order insensitive to small fluctuations in peak lattice depth. However, beyond a certain threshold, the contrast begins to degrade significantly (see Fig.~\ref{fig: Robustness_against_peak_lattice_depth}). To quantify this threshold, we perform exact numerical simulations and plot the contrast as a function of the relative fluctuation $\sigma_R=\Delta \Omega_{R} /\Omega_R$ in Fig.~\ref{fig: Robustness_against_peak_lattice_depth}, where $\Omega_{R}$ denotes the optimal value of peak lattice depth for either beam-splitter or mirror pulse. Each data point corresponds to an average over $10$ realizations with peak lattice depths sampled from a Gaussian distribution $\Omega_{BS|M}\mathcal{N}(1,\sigma_R^2)$ and a fixed momentum width of $\sigma_p=0.05\hbar k_L$ with vanishing COM momentum ($p_0=0$). Theoretical confidence intervals from the five-level model are shown as shaded bands, bounded by $0$ and $1.05 \sigma_R$ deviations from the optimal value. The DS-DBD protocol starts at a high contrast of $97\%$ and remains above $95\%$ up to $\sigma_R=3\%$, outperforming C-DBD, which holds above $95\%$ only up to $\sigma_R=2.5\%$. The OCT strategy exhibits the highest robustness, maintaining over $95\%$ contrast up to $\sigma_R=4.5\%$, outperforming both DS-DBD and C-DBD. The CD-DBD protocol shows the highest sensitivity to lattice depth fluctuations and is therefore omitted from Fig.~\ref{fig: Robustness_against_peak_lattice_depth} for clarity. 
\section{Conclusion}
In summary, we have proposed and analyzed high-contrast Mach–Zehnder atom interferometers based on double Bragg diffraction (DBD) operated under external acceleration. By introducing a tri-frequency laser configuration that compensates differential Doppler shifts, we enable efficient and symmetric beam-splitter and mirror operations even in non-degenerate geometries. To mitigate contrast loss from realistic experimental imperfections—including momentum spread, center-of-mass motion, polarization errors, and lattice-depth fluctuations—we systematically compared four detuning-control strategies: C-DBD, CD-DBD, DS-DBD, and OCT-DBD. 
Our analysis demonstrates that properly engineered detuning-control strategies can elevate double Bragg diffraction interferometers to performance levels previously accessible only with Raman-based schemes. In particular, the OCT strategy simultaneously mitigates Doppler shifts, AC-Stark effects, and polarization errors, enabling robust beam-splitter and mirror operations with near-unity efficiency. Results obtained from the five-level S-matrix theory, validated by exact numerical simulations, show that contrasts exceeding $95\%$ can be maintained across realistic momentum widths and laser-power fluctuations—effectively removing a long-standing limitation of DBD interferometry. These findings establish an experimentally feasible pathway toward high-contrast, large-momentum-transfer atom interferometers. By combining these advanced detuning-controlled DBD pulses with subsequent Bloch oscillations—using the experimentally demonstrated BO efficiency of $99.93\%$ per recoil from Ref.~\cite{Gebbe-NatCom-2021}—our approach is estimated to yield an overall contrast of roughly $56\%$ for a $408\,\hbar k_L$ LMT interferometer. This represents a substantial contrast improvement over state-of-the-art DBD implementations and highlights the potential of detuning-controlled DBD for future high-precision quantum sensors, with applications ranging from terrestrial gravimetry to space-borne tests of fundamental physics.
\section*{Supplementary Material}
The online Supplementary Material contains three sections. \textbf{Sec.~S1} provides a detailed derivation of tri-frequency DBD Hamiltonian (Eq.~\eqref{Eq: H_lab}) in the laboratory frame. \textbf{Sec.~S2} provides a detailed derivation of the unitary transformation from the laboratory-frame DBD Hamiltonian (Eq.~\eqref{Eq: H_lab}) to the twin-lattice center-of-mass frame (Eq.~\eqref{Eq: H_lattice}). \textbf{Sec.~S3} defines the beam-splitter and mirror pulse efficiencies, evaluates their robustness under experimental imperfections, and outlines the cost function used in the OCT optimization.
\section*{acknowledgement}
We thank E. M. Rasel and S. Abend for the insightful discussions and thoughtful comments about the experimental feasibility. The authors gratefully acknowledge financial support from the Deutsche Forschungsgemeinschaft (DFG, German Research Foundation) CRC 1227 274200144 (DQ-mat) within Project A05, Germany's Excellence Strategy EXC-2123 QuantumFrontiers 390837967, and through the QuantERA 2021 co-funded Project No. 499225223 (SQUEIS). We also thank the German Space Agency (DLR) for funds provided by the German Federal Ministry for Economic Affairs and Climate Action (BMWK) due to an enactment of the German Bundestag under Grants No. 50WM2450A (QUANTUS-VI), No. 50WM2253A (AI-Quadrat), and No. 50NA2106 (QGYRO+). R.L. acknowledges the usage of LUH's computer cluster funded by the DFG via Project No. INST 187/742-1 FUGG. N.G. and K.H. acknowledge funding by the AGAPES project - Grant No. 530096754 within the ANR-DFG 2023 Programme.
\section*{Conflict of interest}
The authors have no conflicts to disclose.
\section*{Data availability}
The data that supports the findings of this study are available from the corresponding author upon reasonable request.
\bibliography{references}
\end{document}